\documentclass{article}

\usepackage{graphicx}
\usepackage{makeidx}
\usepackage{multirow}
\usepackage{amssymb}
\usepackage{amsfonts}
\usepackage{amsmath}
\usepackage{amsthm}
\usepackage{natbib}

\usepackage{graphicx}
\usepackage{dcolumn}
\usepackage{bm}

\usepackage[utf8]{inputenc}
\usepackage[T1]{fontenc}
\usepackage{mathptmx}
\usepackage{amsfonts}
\usepackage{bbm}
\usepackage[bottom]{footmisc}

\begin{document}

\title
{On probabilities in quantum mechanics}

\author{Inge S. Helland\\Department of Mathematics, University of Oslo\\P.O. Box 1053, 0316 Oslo, Norway\\ingeh@math.uio.no}

\date{}

\maketitle

\begin{abstract}
This is an attempt to clarify certain concepts related to a debate on the interpretation of quantum mechanics, a debate between Andrei Khrennikov on the one side and Blake Stacey and R$\mathrm{\ddot{u}}$diger Schack on the other side. Central to this debate is the notion of quantum probabilities. I first take up the probability concept in the QBist school, as seen from my point of view, and then give my own arguments for the Born formula for calculating quantum probabilities. In that connection I also sketch some consequences of my approach towards the foundation and interpretation of quantum theory. I discuss my general views on QBism as a possible  alternative interpretation of quantum mechanics before I give some final remarks.

\end{abstract}

\section{Introduction}

The current discussions on the foundation and interpretation of quantum mechanics may be rather intense. Quantum probabilities, as calculated by the Born formula, are central in many of these discussions. As a background for this, it may be useful to look at the various derivations of the Born formula; see Campanella et al. (2020) for some references. My own derivation is given in Helland (2021), and is related to my approach towards quantum foundation, which now has reached its assumed final form in Helland (2024a). This derivation of the Born rule will be repeated below.

This approach to quantum mechanics is a completely new one. It should be looked upon as rather independent of the history of the field, and it is given in a series of articles and books referred to in the reference list below. For the complete mathematical details, I refer again to Helland (2024a). In Helland (2024b) the theory is formulated in an axiomatic way, without giving all the mathematics, and a large set of consequences of the theory is given.

In my opinion, the foundation of quantum mechanics and its interpretation should be sharply linked together. This link was first discussed in Helland (2019).

The present article takes up a special aspect of my theory: My views on quantum probabilities, and a derivation of the Born rule from a set of assumptions.

The Born formula is the central formula in quantum mechanics. It gives the basis for calculation of quantum probabilities.

I will first argue that the simplest version of the Born formula for probabilities in quantum mechanics holds under the following conditions:

1) There is a fixed (physical) context.

2) We have two really different discrete related maximal accessible variables $\theta^a$ and $\theta^b$ in this context, and seek the probability distribution for $\theta^b$, given some value of $\theta^a$, that is, a pure state involving $\theta^a$.

3) The likelihood principle of statistics holds.

4) There exist an inaccessible variable $\phi$ related to the mind(s) of the relevant observer(s) $A$ with the following properties: a) $\theta^a$ and $\theta^b$ are functions of $\phi$; b) As a model, $\phi$ can be imagined to be accessible to the mind of a real or imagined being, seen by $A$ as being superior and perfectly rational in relation to any question involving the variables $\theta^a$ and $\theta^b$.
\bigskip

From these assumptions, the following formula will be derived below, partly following the derivation in Helland (2021).
\begin{equation}
P(\theta^b =v_k^b |\theta^a =u_j^a)=|\langle\psi_j^a |\psi_k^b\rangle |^2 .
\label{Born1}
\end{equation}
Here, $|\psi_j^a\rangle$ is the state vector associated with the event $\theta^a =u_j^a$, and $|\psi_k^b\rangle$ is the state vector associated with the event $\theta^b =v_k^b$. From this simple version of the Born rule, other versions can be derived under natural assumptions.

The argument for (\ref{Born1}) goes in several steps. The basic notion is that of a theoretical variable, which may be a physical variable, but is also assumed to exist in the mind of an observer, or in the joint minds of a group of communicating observers. Theoretical variables may be accessible, possible to measure accurately, or inaccessible. From a mathematical point of view, I only assume the following: If $\lambda$ if a theoretical variable, and $\theta=f(\lambda)$ for some function $f$, then $\theta$ is a theoretical variable. And if $\lambda$ is accessible, then $\theta$ is accessible. (In my book Helland (2021) the theoretical variables were called epistemic variables or e-variables; in some of my articles, I have used the term conceptual variables. I apologize for this confusion.)

Define the following partial ordering among the theoretical variables, and also among the accessible theoretical variables:  
$\theta$ is said to be `less than or equal to' $\lambda$ if $\theta=f(\lambda)$ for some function $f$. A basic assumption behind my theory is: \emph{There exists an inaccessible variable $\phi$ such that all the accessible variables considered can be seen as functions of $\phi$.} This assumption can be easily motivated in simple physical situations.

Using such assumptions together with some specific symmetry assumptions, essentially, the complete quantum formalism is derived in Helland (2024a). It turns out that in the finite-dimensional case, the symmetry assumptions can be dropped, so the Hilbert space theory follows from simple assumptions, the basic one being that there in the given context exist two really different maximal accessible (complementary) theoretical variables. The maximal variables are said to be really different if they are not bijective functions of each other.

The theory in Helland (2024a) in the discrete case can be summurized as follows: Make the assumptions above. Then there exists a Hilbert space $\mathcal{H}$, and all the accessible theoretical variables $\theta$ in the situation have Hermitian operators $A^\theta$ associated with them. The eigenvalues of $A^\theta$ are the possible values of $\theta$. The accessible variable $\theta$ is maximal as such if and only if all eigenvalues are simple. Statevectors can be seen as eigenvectors of some physically meaningful operator. In this sense my axioms (see also Helland, 2024b) imply most elements of the traditional formal quantum theory.

What is left to prove after this, is the Born formula and the Schr\"{o}dinger equation. The first issue will be approached here, for the second, some arguments are given in Helland (2021).

As a special application of the partial ordering defined above, all accessible variables are dominated by $\phi$. Therefore, using Zorn's Lemma, maximal accessible variables with respect to the partial ordering always exist, variables that are just accessible. Physical examples can easily be given, for instance the spin component of a particle in some given direction. People who do not believe in Zorn's Lemma, which is equivalent to the axiom of choice, may take the existence of maximal accessible variables as a separate axiom.

It should be admitted that all this gives a special version of quantum theory: All pure state vectors are assumed to be eigenvectors of some meaningful physical operator. This implies a restriction of the superposition principle, but one can show that certain entangled states are permitted. On the positive side, this restriction leads to simple discussions of several so-called quantum paradoxes, and the whole approach also seems to give connections to aspects of relativity theory and of quantum field theory, see Helland (2023b) and Helland and Parthaserathy (2024).

As a referee pointed out, my approach is also related to the large field of quantum-like models in areas outside physics. In this connection one should first mention the important work of Andrei Khrennikov and collaborators, see Khrennikov (2023), Haven and Khrennikov (2013, 2016), Ozawa and Khrennikov (2021) and the article collectiion Veloz et al. (2023), where further references can be found. Much of this is related to quantum cognition and quantum decision theory, where there is a large recent literature, see Pothos and Busemeyer (2022) and references there.

In Ozawa and Khrennikov (2021) the notion of quantum instrument is defined, and applied to study the order effect in a particular quantum-like model. In the finite-dimensional case, an instrument may be defined trough a set of matrices $\{M_{xi}\}$ satisfying $\sum_{xi} M_{xi}^\dagger M_{xi}=I$, where $x$ denotes the observed outcome of a measurement, and this gives a more general version of the Born formula. This notion is also studied in other articles, including Barndorff-Nielsen et al. (2003), where it implies interesting connection to statistical inference. In the present article I want to limit myself to the simplest possible versions of the Born rule, so quantum instruments are not introduced.

The motivation for writing (a shorter version of) the present article, was a debate between Khrennikov (2024) and two physicists that defined themself as QBists, arguing for a special quantum interpretation of quantum probabilities called QBism, founded by Chris Fuchs; see Caves et al. (2002) and DeBrota et al. (2020a,b,2021). Briefly, the QBists, at least in important part of the literature, regard themselves as subjective Bayesians, a philosophy which in the statistical literature is associated with de Finetti (1972) and Savage (1972). There are very many variants of Bayesianism; as discussed by for instance Good (1983) and von Mises (1981). They are all related to Bayes' formula, which involves prior probabilities, but the basis of these priors is discussed by many authors. A modern account of Bayesianism is given by Bernardo and Smith (2009). Very few statisticians are complete subjective Bayesians today. For a diametric opposite foundation of statistical inference, not involving priors at all, see Schweder and Hjort (2002).

Taking a Theorem by Ozawa (2019) as his point of departure, Khrennikov (2024) recently criticised the QBism interpretation of quantum mechanics. Based on an earlier version of Khrennikov's paper, this critque has been countered by Stacey (2023) and Schack (2023). The discussion is centered around the probability concept. One main purpose of this article is to look at this discussion as seen from my own standpoint. I should also mention that this discussion is taken up from a slightly different point of view in Zwirn (2024).

\section{Quantum probablities according to the QBists}

As a new interpretation of quantum theory, QBism was founded by Chris Fuchs more than 10 years ago, and has since then been developed by a number of authors. Some of the most important articles on the theory are given in the references of Stacey (2023). As in that article, I will not go into details of the theory, but concentrate on the probability concept. Here is a citation from op. cit.:

`According to this school of thought, a probability for an event is nothing more or less than a gambling commitment, a valuation by a specific agent of how much that agent would stake on that event occurring.'

He continues by referring to Khennikov's introduction of certain mathematical entities to describe the situation where two remote observers measure the same variable: Operators $A$, $M_1$ and $M_2$, a state vector $|\psi\rangle$ for the system and another $|\xi\rangle$ for the environment (Khennikov uses $|\xi_1\rangle$ and $|\xi_2\rangle$ for the states of the measurement apparata), plus a one-parameter famlly of unitaries $U(t)$ to represent time evolution.

`Any probability extracted from combining these quantities is necessarily, just like any other probability in personalist Bayesianism, the possession of the agent who commits to it. So there is no way to mix the ingredients $A$, $M_1$, $M_2$ and so forth to arrive at a conclusion that the personal experiences of two agents will always agree, or that they will always disagree, or anything in between.'

I will approach these statement, and any other statements made by physicists, from the point of view of a statistician. The whole science of statistics is built upon probabilies. Bayesianism is one school within statistics, but there are also other schools. As mentioned in the introduction, there are many versions of Bayesianism. The QBists rely on a personalist or subjective version.

As a side remark, statistics is a science that can be explained to intelligent people using fairly everyday terms. One of my own goals is that some day we will be able to do the same with quantum physics. Helland (2024b) is a beginning.

So concentrate first on `a probability for an event is nothing more or less than a gambling commitment'. Taken to its extreme, QBists seem to think in terms of gambling all the time; every time we make a decision we make an internal bet. I will claim that ordinary people neither think nor act in this way. We go through life making decision after decision, and very rarely we think in terms of gambling when we make these decisions. In my opinion, so also when we make statements of probabilities of events.

As a referee points out, QBists talk about gambling in some generalized sense, not necessary expressible in monetary terms. This is true. I agree with much of the QBist philosophy, but not all. I particuler, we agree that a quantum state is a state of knowledge. In my theory, this knowledge is attached to an observer/agent or to a communicating group of agents.

A major difference between my views and the QBist views on probability, see Caves et al. (2002), is that the QBists rely on a Dutch book argument connected to the agent himself, while I only assume the the agent has ideals that he think of as rational, as made precise by a Dutch book argument.

Classically, the concept of probability may have many foundations. Probabilities may be based upon symmetries, like when throwing a die or evaluating an opinion poll, it can be based upon subjective judgement, or it can be based on much data and long experience, like when a meteorologist makes a probability statement. Only in the subjective case, it is possible at all to talk about some internal gambling procedure. I will claim that even in that case, people tend to make probability statements without having any bets in mind.

In Caves et al. (2002), arguments for quantum probabilities are given by using a Dutch book argument for an agent, de Finetti's representation theorem for mixed quantum states, and Gleason's theorem. The weak point, as I see it, is the assumption that the agent is rational, as expressed by a Dutch book argument. 

We are in no way always rational when making our decisions. The decisions are often made by using an intuition that has been formed by a long history, based on experinces and contact with other people. During this process, we have our limitations, as described from a quantum theory background in Helland (2022b, 2023d).

So, in my view, even in the case of quantum probabilities, the QBist probability concept is unsatisfactory. For quantum probabilities, we have to take a closer look at the background for the Born theorem. The QBists' view on the Born rule will be discussed later. My own background for this rule is described below,

So, in a number of Sections I will now give my arguments. They are based upon my point of view as a statistician. One basic goal of my research has been to try to build a bridge between the statistical culture and the quantum mechanical culture, and a part of this goal has been to find a foundation of quantum theory that can be explained to scientists like statisticians. This is very difficult with the existing formalism, but I think it is easier using my alternative foundation.

I will come back to the QBist views of probabilities in Sections 8 and 9.

\section{The likelihood principle}

The basis for nearly all statistical inference is a statistical model, a probability model for the data $z$, given some defined parameter $\theta$. In most cases, both $z$ and $\theta$ are multivariate, i.e., can be seen as vectors. The model is then given by a probability function for the data, given the parameter, $p(z|\theta)$. In the continuous case, $p$ is a probability density, in the discrete case a point probability.

The likrlihood of the data is defined as $L(\theta |z)=p(z|\theta)$, the probability function seen as a function of the parameter.

Statistical inference builds upon several principles. One of these is the likelihood principle.. The principle says roughly that all relevant information in some experiment is contained in this likelihood. The principle can be derived from other principles; see Helland (2021), but it can also be argued for independently.
A version of the likelihood principle will be used below as a part of the motivation behind Born's formula.
\bigskip

\textbf{The Generalized Likelihood Principle}
\textit{Consider two experiments with equivalent contexts $\tau$, and assume that $\theta$ is the same
full parameter in both experiments. Suppose that two observations $z_1 ^*$ and $z_2 ^*$ have proportional likelihoods in the two experiments, where the
proportionality constant $c$ is independent of $\theta$. Then these two observations produce the same experimental evidence on $\theta$ in this context.}
\bigskip

The term `experimental evidence' is here left undefined, and can be specified in any desirable direction; see a closer discussion in Helland (2021).

Two contexts are said to be equivalent if one can establish a one-to-one function between all variables involved. It is important for my development that the context is kept fixed. This aspect makes the generalized likelihood principle weaker than the principle as formulated
in the literature, in particular in Berger and Wolpert (1988). Paradoxes like what the ordinary likelihood principle seems to imply in the
following situation are avoided.
\bigskip

\textit{Example} Suppose that $s_1 , s_2 ,\ldots$ are independent, identically (iid) distributed variables with $P(s=1)=\theta$ and $P(s=0)=1-\theta$, in statistical language, iid Bernoulli
variables with parameter $\theta$. In experiment $\mathcal{E}_1$, a fixed sample size of
ten observations is decided upon, and the important summary observation (in statistical language the sufficient statistics) $t_1 =\sum_{i=1} ^{10} s_i$ turns out to be $t_1=8$. In experiment $\mathcal{E}_2$, it is decided to
take observations until a total of 2 zeroes has been observed. Then assume that the sufficient statistics $t_2 =\sum s_i$ also turns out to take the value 8. The two likelihoods are
proportional, but the contexts are different, so the intuition that the two experiments may lead to different inference on $\theta$ is supported by my version of the
likelihood principle. For further discussion of this example, see Berger and Wolpert (1988) and references there.
\bigskip

The introduction of a context makes my formulation of the likelihood principle far less controversial than the ordinary formulation. According to the ordinary
principle, the way data are obtained is irrelevant to inference; all information is contained in the likelihood. Thus sampling plans, randomization procedures,
and stopping rules are irrelevant according to a common interpretation of the ordinary principle. Furthermore, common frequentist concepts like bias,
confidence coefficients, levels and powers of statistical tests, etc., are irrelevant, as they depend on the sample space, not only on the observed observations. In
my formulation, all these concepts are related to the context. Also Bayesian priors, if needed, are contained in the context. Maximum likelihood estimation
can not be derived from the likelihood principle, but is obviously permissible as a method of obtaining reasonable proposals for estimates in general.

An important special case of the generalized likelihood principle is when the proportionality constant $c$ is equal to 1. Then the two observations $z_{1}^*$
and $z_{2}^*$ are assumed to have equal likelihoods. Again an important special case is when the two experiments are identical. A consequence of the generalized likelihood
principle is then that \emph{all experimental evidence, given the context, is a function of the likelihood of the experiment, i.e, is contained in the likelihood function}.

From this point of view the situation is similar in quantum mechanics as in in ordinary statistics. Here, in a given situation, we may have a model for the data $z$ depending upon the context $\tau$ and the theoretical variable of interest $\theta$, expressed by a point probability or probability density $p(z|\tau,\theta)$.  Thus, even though $\theta$ may be discrete, from a statistical point of view it acts as a parameter in the model. An eventual extra parameter $\eta$ in such a model will be assumed known from earlier experiments of the same type, and may be included in the context. This gives a unique likelihood $L(\theta |z,\tau )=p(z|\tau ,\theta )$. And also in this situation the relevant discussion in Helland (2021) seems to imply that the generalized likelihood principle above holds true.

\section*{The focused likelihood principle}

In this section I assume a discrete quantum formulation. One can for instance think of a spin component in a fixed direction to be determined. I assume a measurement situation where the data contains some noise, hence a likelihood for the discrete parameter $\theta$, given data $z$ as $ L(\theta|z)=p(z|\theta)$, where $p$ is the probability density or point probability of the data.

Assume now that the quantum mechanical system is prepared in some state and that we want to do an experiment related to the unknown theoretical variable $\theta^b$. Given then the focused question $b$, the  theoretical variable
$\theta^b$ plays the role similar to a parameter in statistical inference. Inference can be done by preparing many independent units in the same state. Inference is then made from data $z^b$. All inference theory that one finds in standard statistical  texts like Lehmann and Casella (1998) applies. In particular, the concepts of
unbiasedness, equivariance, average risk optimality, minimaxity and admissibility apply. None of these concepts are much discussed in the physical literature,
first because measurements there are often considered as perfect, at least in elementary texts, secondly because, when measurements are considered in the
physical literature, they are mostly discussed in other terms. 

Whatever kind of inference we make on $\theta^b$, we can take as a point of departure the statistical model and the generalized likelihood principle of the previous Section. Hence after an experiment is done, and given some context $\tau$, all evidence on $\theta^b$ is contained in the likelihood
$p(z^b|\tau ,\theta^b )$, where $z^b$ is the data relevant for inference on $\theta^b$, also assumed discrete. This is summarized in the
\emph{likelihood effect}:
\begin{equation}
F^b(\bm{u}^b ; z^b ,\tau)=\sum_j p(z^b |\tau ,\theta^b =u^b_j ) |b;j\rangle\langle b;j|,
\label{xx}
\end{equation}
where the pure state $|b;j\rangle$ corresponds to the event $\theta^b = u^b_j$. 
\bigskip

Interpretation of the likelihood effect $F^b(z^b ,\tau)$:

(1) We have posed some inference question on the accessible theoretical variable.
 $\theta^b$. (2) We have specified the relevant likelihood for the data. The question itself and the likelihood for all possible answers of the question, formulated in terms of state
vectors, can be recovered from the likelihood effect.

The likelihood effect is closely connected to the concept of an operator-valued measure POVM; see a discussion in Helland (2021). Since the focused question assumes discrete data, each likelihood is in the range $0\le p \le 1$. In the quantum mechanical literature, an effect is any operator
with eigenvalues in the range $[0,1]$.

Some qualifications must be made relative to the above interpretation, however, if we want to be precise. We have the freedom to redefine the theoretical variable in the case of coinciding eigenvalues in the likelihood effect, that is, if
$p(z^b |\tau ,\theta^b =u_j^b)=p(z^b |\tau ,\theta^b =u_l ^b)$ for some $j$, $l$. An extreme case is the likelihood effect $F(\boldsymbol{u}^b ; z^b , \tau)=I$,
where all the likelihoods are 1, that is, the probability of $z$ is 1 under any considered model. One could have defined the likelihood effect from appropriate eigenvalue spaces, but for the following mathematical result, the definition (\ref{xx}) is convenient.

We have the following result on the likelihood effects:
\bigskip

\textbf{Proposition 1} \textit{Let two experiments $b$ and $c$ be given together with two data points $z^b$ and $z^c$ of these experiments. Assume that $b$ and $c$ are  such that}\vspace*{-3pt}
\begin{equation}
F^b(\bm{u}^b;z^b,\tau)= F^c(\bm{u}^c;z^c,\tau).
\label{effect}\vspace*{-3pt}
\end{equation}
\textit{Then we can order the states such that}

(1) \textit{$p(z^b |\tau ,\theta^b =u^b_j )= p(z^c |\tau ,\theta^c =u^c_j )$ for each $j$.}

(2) \textit{Introduce the class of indices $C_i$ such that $p(z^b |\tau ,\theta^b =u_k^b)=p(z^b |\tau ,\theta^b =u_l ^b)$ whenever $k,l\in C_i$, and these likelihoods are different when $k$ and $l$ belong to different $C_i$-classes, similarly $D_i$ for $p(z^c |\tau ,\theta^c =u^c_k )$. Then we have}\vspace*{-3pt}
\[\sum_{k\in C_i} |b;k\rangle\langle b;k| = \sum_{k\in D_i} |c;k\rangle\langle c;k|\vspace*{-3pt}\]
\textit{for all $i$.}

\textit{On the other hand, if (1) and (2) are satisfied, then (\ref{effect}) holds.}
\bigskip

The last part is fairly trivial. The direct part is proved in Appendix 1.

Return now to the generalized likelihood principle of the previous Section. Recall that this principle is fairly reasonable in our setting, where we condition upon the context $\tau$. In statistics, the likelihood principle says the
following: If two experiments have proportional likelihood, with constant of proportionality independent of the parameter, they produce the same experimental
evidence about the parameter. Here experimental evidence is left undefined. In my approach towards quantum mechanics, where one focuses on a specific question, we must in addition demand that this focused question is the same; that is, the set of corresponding projections must be the same.

The following principle follows:
\bigskip

\textbf{The Focused Generalized Likelihood Principle (FGLP)}
 \textit{Consider two potential experiments $b$ and $c$ in some setting with equivalent contexts $\tau$, and assume that the inaccessible theoretical variable $\phi$ is the same in both experiments. Suppose that the two observations $z_1^b$ and $z_2^c$ have equal likelihood effects in the two experiments.}

 \textit{Then}

 \textit{(A) The questions posed in the two experiments are equivalent in the sense that one can use the same Hilbert space $\mathcal{H}$ to describe the results of the experiments, and that the corresponding set of eigenvector spaces (i.e., the orthogonal resolution of the identity) are equal. This implies a one-to-one relation between the theoretical variables $\theta^b$ and $\theta^c$.}

 \textit{(B) The two observations produce equivalent experimental evidence on the relevant theoretical variables in this context and given this question.}
 \bigskip

\textbf{Proposition 2}
\textit{The focused generalized likelihood principle follows from the generalized likelihood principle.}
\bigskip

\begin{proof}
The Proposition is trivial if we know that the theoretical variables are the same in the two experiments. If not, we have a situation where Eq. (\ref{effect}) holds. Then $\theta^b$-operator$=\sum u_k^b \Pi_k^b$, $\theta^c$-operator$=\sum u_k^c \Pi_k^c$, where the equality of the projection operators after a suitable ordering follows from Proposition 1, (2). The eigenvalues $u_k^b$ are all different, similarly the eigenvalues $u_k^c$. These are the answers to the questions connected to $\theta^b$ and $\theta^c$; the questions are equivalent since the set of projection operators coincide. The conclusion (B) follows from Proposition 1, (1) and the ordinary (generalized) likelihood principle.
\end{proof}

\underline{Remark:} Strictly speaking, this proof uses the assumption that $p(z^b|\tau, \theta^b=u_k^b) = p(z^b |\tau, \theta^b=u_j^b)$ implies $u_k^b=u_j^b$, and similarly for $z^c$. This is in agreement with the arbitrariness of $\theta^b$ in relation to $F^b$ discussed above. Below, in the proof of Born's formula, I will use FGLP in the very special case of a perfect experiments, where observations can be taken to be equal to parameter values, and there this assumption is trivial.

Two contexts are considered equivalent if they are one-to-one functions of each other. \emph{The principle FGLP  says that both the question posed and the experimental evidence are functions of the likelihood effect and the context of the experiment.}

\section{Rationality and experimental evidence}

Throughout this Section and the next one, I will consider a fixed context $\tau$ and a fixed epistemic setting in this context. The inaccessible theoretical variable is $\phi$, and I
assume that the accessible theoretical variables $\theta^b$ take a discrete set of values.  Let the data behind the potential experiment connected to $\theta^b$ be $z^b$, also assumed to
take a discrete set of values. I will assume throughout this Section and the next one that an experimentalist $B$ is in such a context partly determined by the fact that he previously has performed an perfect experiment connected to a maximally accessible theoretical variable $\theta^a$ and obtained the answer $\theta^a=u_k$, so that his state can be described by a Hilbert space $\mathcal{H}$ and a vector $|a;k\rangle$ in $\mathcal{H}$. For the background for such assumption, see Helland (2024a,b), where the basic condition is that there, in connection to an agent or to a communicating group of agents in a given context, exist two, really different maximal accessible (complementary) variables.

So let a single agent $B$ be in this situation, and let all theoretical variables be attached to $B$, although he also has the possibility to receiving
information from others through part of the context $\tau$. He has the choice of doing different experiments $b$, and he also has the choice of choosing different
models for his experiment through his likelihood $p_{B}(z^b |\tau ,\theta^b)$. The experiment and the model, hence the likelihood, should be chosen before
the data are obtained. All these choices are summarized in the likelihood effect $F^b$, a function of the at present unknown data $z^b$. For use after the experiment,
he should also choose a good estimator  $\widehat{\theta^b}$, and he may also have to choose some loss function, but the principles behind these latter choices will be considered as part of the context $\tau$.

If $B$ chooses to do a Bayesian analysis, the estimator should be based on a prior
$\pi_B (\theta^b |\tau)$. We assume that he is trying to be as rational as possible in all his choices, and that this rationality is connected to his loss function or to
other criteria. What should be meant by experimental evidence, and how should it be measured? As a natural choice then, let the experimental evidence that we are seeking, be the posterior probability  for some fixed value of $\theta^b$, given the data. From the agent $B$'s point of view this is given by:

\[q=p_B^b(\theta^b=u_j^b |z^b,\tau )=\frac{p_B (z^b |\tau , \theta^b =u_j^b )\pi_B (\theta^b=u_j^b |\tau)}{\sum_i p_B^b (z^b |\tau , \theta^b =u_i^b )\pi_B (\theta^b=u_i^b |\tau)}, \]
assuming the likelihood chosen by $B$ and $B$'s prior $\pi_B$ for $\theta^b$.

Some Bayesians claim that their own philosophy is the only one which is
consistent with the likelihood principle. For my own view on this, see Helland (2021). In a non-Bayesian analysis, we can let the concept of experimental evidence be tied to the confidence distribution, given the context, see Schweder and Hjort (2002).  (Another non-Bayesian analysis is fiducial inference; see Hannig et al., 2016). This may also give a conclusion in terms of a probability, in this case an epistemic probability, a concept that is further discussed in Helland (2021). Also in such an analysis we must assume $B$ to be as rational as possible.

In any case we fix $b$ and $j$ from now on, and take $q=q_j^b = P(\theta^b = u_j^b |\mathrm{data})$, an epistemic probability.

I have to make precise in some way what is meant by the rationality of the experimentalist $B$. He has to make many difficult choices on the basis of uncertain
knowledge. His actions can partly be based on intuition, partly on experience from similar situations, partly on a common scientific culture and partly on advices
from other persons. These other persons will in turn have their intuition, their experience and their scientific education. Often $B$ will have certain explicitly formulated
principles on which to base his decisions, but sometimes he may have to dispense with some of the principles. In the latter case, he has to rely on some `inner voice', a conviction which
tells him what to do.

So in the case where $B$ can not himself be seen as a perfectly rational Bayesian, a case that I will concentrate on below, I will formalize this by introducing a perfectly rational superior actor $D$, to which all these principles, experiences and convictions can be related. I will assume that $B$ in his actions is inspired by $D$, so in this sense, $D$ has some influence on $B$'s decisions. 
I may assume that $D$ has priors, so that he can do a Bayesian analysis. These priors can be based on symmetry considerations, but also on other considerations. The
 experimental evidence will then be defined as \emph{the aposteriori probability of the variable} $\theta^b$ \emph{from} $D$'\emph{s point of view}, say the probability $q$ that $\theta^b$ takes some fixed value $u_j^b$, given the data. By the FGLP this must again be a function of the likelihood effect $F^b$.
\begin{equation}
B\text{'s\ experimental\ evidence,\ related\ to\ D}=q(F^b(\boldsymbol{u}^b ;z^b ,\tau))
\label{evidence}
\end{equation}
under the assumption that the experiment connected to some variable $\theta^b$ is to be done. 

Alternatively, I may also asume that $D$ is a frequentist, and that he has epistemic probabilities connected to $\theta^b$ as found from a confidence distribution; see Schweder and Hjort (2002).  Again, by the focused likelihood principle, these epistemic probabilities must be of the form (\ref{evidence}).

In any case, for the derivation of Born's formula below, I will make one more crucial assumption: $q$ as defined here gives the real probability that $\theta^b$ takes the fixed value $u_j^b$ in the given context.

The superior actor $D$ represents the scientific ideals of the experimentalist $B$, and my main point is that $D$ should be perfectly rational,

In this article I have not tried to develop a theory of decisions. In Helland (2023a,c) I have argued that there is a close connection between the foundation of quantum mechanics and quantum decision theory.  Here one must be careful, however. Quantum decision theory takes its departure in the ordinary mathematical formulation of quantum theory, in particular in Born's rule for calculating probabilities. Using this theory here, where I am preparing to derive Born's rule, will lead to circular reasoning. Instead I will now take \emph{decision} as a primitive concept.

An important point is that decisions made by our minds are not the same as straightforward computerlike calculations. Human decisions are based on the functioning of and the interplay between conscious and subconscious processes in the brain. 

As said, in a scientific connection we assume that $D$ is perfectly rational. This can be formalized mathematically by considering a hypothetical betting situation for $D$ against a bookie, nature
$N$. A similar discussion was recently done using a more abstract language by Hammond (2011). But note: I do not see any human scientist, including myself, as being perfectly rational in all situations. We can try to be as rational as possible, but we have to rely on some underlyng rational ideals that partly determine our actions.

So let the hypothetical odds of a given bet for $D$ be $(1-q)/q$ to 1, where $q$ is the probability as defined by (\ref{evidence}). This odds specification is a
way to make precise that, given the context $\tau$ and given the question $b$, the bettor's probability that the experimental result takes some value, say $u_j^b$, is given by
$q$: For a given utility measured by $x$, the bettor $D$ pays in an amount $qx$ --- the stake --- to the bookie.  After the experiment the bookie pays out an amount
$x$ --- the payoff --- to the bettor if the result of the experiment takes the value $\theta^b=u_j^b$, otherwise nothing is paid.

The rationality of $D$ is formulated in terms of
\bigskip

\textbf{The Dutch Book Principle}
\textit{No choice of payoffs in a series of bets shall lead to a sure loss for the bettor.}
\bigskip

 For a related use of the same principle, see Caves et al. (2002). It is very important that the principle is related to a fixed context; the hypothetical superior bettor $D$ is also bounded by this context. This has also consequences for my derivation of Born's formula below.
 
 It is also important that this whole discussion is limited to a context where the observer $B$ during his decision has just two related maximal variables in his mind. The superior, hypothetical bettor $D$ must then also be related to such a context. This is important in the derivation of Born's formula below, and it is also important when I later generalize to macroscopic decisions. Several recent articles discuss quantum cognition as modeled by quantum probabilities; for a recent review, see Pothos and Busemeyer (2022). In my opinion, such models could be based upon a derivation of Born's formula and a simple model for a person's decisions. But this model should then be limited to decisions between two related maximal variables. The superior, hypothetical actor $D$ must also be seen in this light.
\bigskip

\textbf{Assumption 1}\label{ch5:ass2}
\textit{Consider in the context $\tau$ an epistemic setting where the FGLP is satisfied, and the whole
situation is observed by an experimentalist $B$ whose decisions are influenced by a superior actor $D$ as described above. Assume that $D$'s probabilities $q$ given by (\ref{evidence}) are taken as the
experimental evidence, and that $D$ can be seen to be rational in agreement with the Dutch book principle.}
\bigskip

A situation where Assumption 1  holds will be called a \textit{rational epistemic setting}. It will be assumed to
be implied by essential situations of quantum mechanics. Later I will discuss whether or not it also can be coupled to certain macroscopic situations.
\bigskip

\textbf{Theorem 1} \textit{Assume a rational epistemic setting, and assume a fixed context $\tau$. Let $F_{1}$ and $F_{2}$ be two likelihood effects in this setting,
and assume that $F_{1}+F_{2}$ also
is an effect. Then the experimental evidences, taken as the epistemic probabilities related to the data of the performed experiments, satisfy}
\[q(F_{1}+F_{2}|\tau)=q(F_{1}|\tau)+q(F_{2}|\tau).\]

\begin{proof}
The result of the theorem is obvious, without making Assumption 1, if $F_{1}$ and $F_{2}$ are likelihood effects connected to experiments on the same
variable $\theta^b$. We will prove it in general. Consider then two experiments 1 and 2, 1 with variable $\theta^b$ and likelihood effect $F_1$, and 2 with theoretical variable $\theta^c$ and likelihood effect $F_2$, and let B choose between these two experiments. Let $z$ be the data of the chosen experiment, and let the corresponding posterior probabilities /confidence probabilities connected to $D$ be $q_1 =P(\theta^b =u_j^b |z,\tau)=q(F_1|\tau)$ and $q_2 =P(\theta^c =u_k^c|z,\tau)=q(F_2|\tau)$ for some $j$ and $k$.

Assume that $B$ does a randomized choice: With an unbiased coin he makes the choice betwen the experiments 1 or 2. The likelihood effect connected to the whole experiment, including the coin toss, is $(F_1+F_2)/2$.

Let $q_0$ be the posterior probability / confidence probability for $D$ connected to this full experiment, including the randomization. One might perhaps argue tentatively at once that $q_0$ must be equal to $(q_1+q_2)/2$, but I will show in detail that this follows from the Dutch Book Principle.

Let $D$ make his bets, one for the experiment 1, one for the experiment 2, and one for the full randomized experiment. Let the corresponding payoffs chosen by Nature be $x_1 , x_2$ and $x_0$. Imagine that this, including the randomization, is repeated a large number of times.

If experiment 1 occurs in the randomization, the payoff for the randomized experiment is replaced by the expected payoff
$x_{0}/2$, similarly if experiment 2 occurs. The net expected amount the bettor receives is then
\[x_{1}+\frac{1}{2}x_{0}-q_{1}x_{1}-q_{2}x_{2}-q_{0}x_{0}=(1-q_{1})x_{1}-q_{2}x_{2}-(1-2q_{0})\frac{1}{2}x_{0}\]
if experiment 1 is done and the conclusion is $\theta^b=u_j^b$,

\[x_{2}+\frac{1}{2}x_{0}-q_{1}x_{1}-q_{2}x_{2}-q_{0}x_{0}=-q_{1}x_{1}-(1-q_{2})x_{2}-(1-2q_{0})\frac{1}{2}x_{0}\]
if experiment 2 is done and the conclusion is $\theta^c =u_k^c$,

\[-q_{1}x_{1}-q_{2}x_{2}-2q_{0}\cdot \frac{1}{2}x_{0}\ \mathrm{otherwise}.\]

This conclusion may be drawn from many repeated experiments.

The payoffs $(x_{1},x_{2},x_{0})$ can be chosen by nature $N$ in such a way that it leads to sure loss for the bettor $D$ if not the determinant of this system is zero:
\[
0= \left| \begin{array}{ccc}
1-q_{1}&-q_{2}&1-2q_{0}\\
-q_{1}&1-q_{2}&1-2q_{0}\\
-q_{1}&-q_{2}&-2q_{0}
\end{array}\right| =q_{1}+q_{2}-2q_{0}.
\]
Thus we must have
\[q(\frac{1}{2}(F_{1}+F_{2})|\tau)=\frac{1}{2}(q(F_{1}|\tau)+q(F_{2}|\tau)).\]
If $F_{1}+F_{2}$ is an effect, the common factor $\frac{1}{2}$ can be removed by changing the likelihoods, and the result follows.
\end{proof}
\bigskip

\textbf{Corollary 1} \textit{Assume a rational epistemic setting in the context $\tau$. Let $F_{1}$, $F_{2}$, \ldots be likelihood effects in this setting, and assume that $F_{1}+F_{2}+\ldots$ also
is an effect. Then}
\[q(F_{1}+F_{2}+\ldots|\tau)=q(F_{1}|\tau)+q(F_{2}|\tau)+\ldots.\]

\begin{proof}
The finite case follows immediately from Theorem 1. Then the infinite case follows from monotone convergence.
\end{proof}

The result of this Section is quite general. In particular the loss function and any other criterion for the success of the experiments are arbitrary. So far I have
assumed that the choice of experiment $b$ is given, which implies that it is the same for $B$ and for $D$. However, the result also applies to the following
different situation: Let $B$ have some definite purpose for his experiment, and to achieve that purpose, he has to choose the question $b$ in a clever manner, as
rationally as he can. Assume that this rationality is formalized through the actor $D$, who has the ideal likelihood effect $F$ and the experimental evidence
$q(F|\tau)$. If two such questions can be chosen, the result of Theorem 1 holds, with essentially the same proof.

\section{The Born formula}\label{ch5:sec6}

\subsection{The basic formula}\label{ch5:sec6.1}

Born's formula is the basis for all probability calculations in quantum mechanics. In textbooks it is usually stated as a separate axiom, but it has also been argued for by
using various sets of assumptions; see Helland (2008) and Campanella et al. (2020) for some references. In fact, the first argument for the Born formula, assuming that there is an affine mapping from set of density functions to the corresponding probability functions, is due to von Neumann (1927); see Busch et al. (2016). In Helland (2006), Helland (2008) and Helland (2010) the formula was proved under rather strong assumptions. Here I will use assumptions which are as weak as possible; I will base the discussion upon the result of the previuos Sections.

I begin with a very elegant recent theorem by Busch (2003). For completeness I reproduce the proof for the
finite-dimensional case in Appendix 2.

Let in general $\mathcal{H}$ be any separable Hilbert space. Recall that an
effect $F$ is any operator on the Hilbert space with eigenvalues in the range $[0,1]$. A generalized probability measure $\mu$ is a function on the effects with the properties
\[
\begin{array}{l}
(1)\ 0\le \mu(F)\le 1\ \text{for\ all}\ F,\\
(2)\ \mu(I)=1,\\
(3)\ \mu(F_{1}+F_{2}+\ldots)=\mu(F_{1})+\mu(F_{2})+\ldots\ \mathrm{whenever}\ F_{1}+F_{2}+\ldots\le I.
\end{array}
\]

\textbf{Theorem 2} (Busch, 2003) \textit{Any generalized probability measure $\mu$ is of the form $\mu(F)=\mathrm{trace}(\rho F)$ for
some density operator $\rho$.}
\bigskip

It is now easy to see that $q(F|\tau)$ on the likelihood effects of the previous Section is a generalized probability measure if Assumption 1 holds:
(1) follows since $q$ is a probability; (2) since $F=I$ implies that the likelihood is 1 for all values of the theoretical variable; finally (3) is a
consequence of the corollary of Theorem 1. Hence there is a density operator $\rho =\rho(\tau)$ such that
$p(z|\tau)=\mathrm{trace}(\rho(\tau)F)$ for all ideal likelihood effects $F=F(z)$. This is a result which is valid for all experiments.

The problem of defining a generalized probability on the set of effects is also discussed in Busch et al. (2016).

Define now a \textit{perfect experiment} as one where the measurement uncertainty can be disregarded. The quantum mechanical literature operates
very much with perfect experiments which result in well-defined states $|j\rangle$.  From the point of view of statistics, if, say the 99\% confidence or
credibility region of $\theta^b$ is the single point $u_{j}^b$, we can infer approximately that a perfect experiment has given the result $\theta^b =u_{j}^b$.

In our epistemic setting then: We have asked the question: `What is the value of the accessible variable $\theta^b$?', and are
interested in finding the probability of the answer $\theta^b =u_{j}^b$ though a perfect experiment. If $u_j^b$ is a non-degenerate eigenvalue of the operator corresponding to $\theta^b$, this is the probability of a well-defined state $|b;j\rangle$.
Assume now that this probability is sought in a setting defined as follows: We have previous knowledge of the answer
$\theta^a =u_{k}^a$ of another maximal question: `What is the value of $\theta^a$?' That is, we know the state $|a;k\rangle$. ($u_k^a$ is non-degenerate.)

These two experiments, the one leading to $|a;k\rangle$ and the one leading to $|b;j\rangle$, are assumed to be performed in equivalent contexts $\tau$.
\bigskip

\textbf{Theorem 3} [Born's formula] \textit{ Assume a rational epistemic setting. In the above situation we have:}
\begin{equation}
P(\theta^b =u_{j}^b |\theta^a =u_{k}^a)=|\langle a;k|b;j\rangle|^2 .
\label{Born}
\end{equation}

\begin{proof}
By the theory in Helland (2024a), both the variable $\theta^a$ and the variable $\theta^b$ have operators with non-degenerate eigenvalues. Fix $j$ and $k$, let $|v\rangle$ be either
$|a;k\rangle$ or $|b;j\rangle$, and consider likelihood effects of the form $F=|v\rangle\langle v|$. This corresponds in
both cases to a perfect measurement of a maximally accessible parameter with a definite result. By Theorem 2 above there exists a density operator
$\rho^{a,k} =\sum_{i}\pi_{i}(\tau^{a,k})|i\rangle\langle i|$ such that $q(F|\tau^{a,k})=\langle v|\rho^{a,k}|v\rangle$, where
$\pi_{i}(\tau^{a,k})$ are non-negative constants adding to 1. Consider first $|v\rangle=|a;k\rangle$. For this case one must have
$\sum_{i} \pi_{i}(\tau^{a,k})|\langle i|a;k\rangle|^2 =1$ and thus $\sum_{i} \pi_{i}(\tau^{a,k}) (1-|\langle i|a;k\rangle|^2)=0$. This implies
for each $i$ that either $\pi_{i}(\tau^{a,k})=0$ or $|\langle i|a;k\rangle|=1$. Since the last condition implies $|i\rangle=|a;k\rangle$ (modulus an
irrelevant phase factor), and this is a condition which can only be true for one $i$, it follows that $\pi_{i}(\tau^{a,k})=0$ for all other $i$ than this one,
and that $\pi_{i}(\tau^{a,k})=1$ for this particular $i$. Summarizing this, we get $\rho^{a,k}=|a;k\rangle\langle a;k|$, and setting
$|v\rangle=|b;j\rangle$, Born's formula follows, since $q(F|\tau^{a,k})$ in this case is equal to the probability of the perfect result $\theta^b =u_{j}^b$.
\end{proof}

\section{Consequences}

Here are three easy consequences of Born's formula:
\begin{enumerate}
\item[1.] If the context of the system is given by the state $|a;k\rangle$, and $A^b$ is the operator corresponding to the variable $\theta^b$, then the expected value of a
perfect measurement of $\theta^b$ is $\langle a;k|A^b |a;k\rangle$.

\item[2.] If the context is given by a density operator $\rho$, and $A$ is the operator corresponding to the variable $\theta$, then the expected value of a
perfect measurement of $\theta$ is $\mathrm{trace}(\rho A)$. 

\item[3.] In the same situation the expected value of a perfect measurement of $f(\theta)$ is $\mathrm{trace}(\rho f(A))$.
\end{enumerate}

\begingroup
\renewcommand{\proofname}{Proof of 1.}
\begin{proof}
\[\mathrm{E}(\theta^b |\theta^a =u_k^a)=\sum_i u_i^b P(\theta^b =u_i^b |\theta^a =u_k^a)\]
\[=\sum_i u_i^b \langle a;k |b;i \rangle \langle b;i |a;k \rangle = \langle a;k|A^b |a;k\rangle .\]
\end{proof}
\endgroup

\begingroup
\renewcommand{\proofname}{Proof of 2.}
\begin{proof}
Let $\rho=\sum_k \pi_k^a |a;k\rangle\langle a;k|$ and $A=\sum_j u_j^b |b;j\rangle\langle b;j|$. Then from 1.
\[\mathrm{E}(\theta)=\sum_k \pi_k^a  \langle a;k|A |a;k\rangle =\mathrm{trace} \sum_k \pi_k^a |a;k\rangle\langle a;k| A. \]
\end{proof}
\endgroup

A consequence of 3. above is that $\theta=\theta^b$ does not need to be maximal in order that a Born formula should be valid; see also below.

As an application of Born's formula, we give the transition probabilities for electron spin. For a given direction $a$, define the
variable $\theta^a$ as +1 if the measured spin component by a perfect measurement for the electron is $+\hbar/2$ in this direction, $\theta^a =-1$ if the
component is $-\hbar/2$. Assume that $a$ and $b$ are two directions in which the spin component can be measured.
\bigskip

\textbf{Proposition 3} \textit{For the qubit spin components we have}\vspace*{-3pt}
\[P(\theta^b =\pm 1|\theta^a =+1)=\frac{1}{2}(1\pm \mathrm{cos}(a\cdot b)).\vspace*{-3pt}\]
\bigskip

This is proved in several textbooks, for instance Holevo (2001), from Born's formula.  A similar proof using the Pauli spin matrices is also given in Helland (2010).

\subsection{Perfect measurements}

Measurements of theoretical variables is discussed in Helland (2021), here I will look at the case of a perfect measurement. Assume that we know the state $|\psi\rangle$ of a system, and that we want to measure a new variable $\theta^b$. This can be discussed by means of the projection operators $\varPi_j^b =|b;j\rangle\langle b;j|$. First observe that by a simple calculation from Born's formula\vspace*{-3pt}
\begin{equation}
P(\theta^b=u_j^b |\psi)= \|\varPi_j^b |\psi\rangle \|^2.
\label{Born21}\vspace*{-3pt}
\end{equation}

It is interesting that Shrapnel et al. (2017)  recently simultaneously derived \emph{both} the Born rule and the well-known collapse rule from a knowledge-based perspective. I say more about the collapse rule in Helland (2021), but in this article I will just assume this derivation as given.
Then, after a perfect measurement $\theta^b=u_j^b$ has been obtained, the state changes to
\[|b;j\rangle=\frac{\varPi_j^b |\psi\rangle}{\|\varPi_j^b |\psi\rangle\|}.\]

Successive measurements are often of interest. We find
\begin{gather}
P(\theta^b=u_j^b\ \text{and\ then}\ \theta^c=u_i^c|\psi)=P(\theta^c=u_i^c |\theta^b=u_j^b)P(\theta^b=u_j^b |\psi)\nonumber\\
=\|\varPi_i^c \frac{\varPi_j^b |\psi\rangle}{\|\varPi_j^b |\psi\rangle\|}\|^2 \|\varPi_j^b |\psi\rangle \|^2= \|\varPi_i^c \varPi_j^b |\psi\rangle \|^2.
\label{successive}
\end{gather}

In the case with multiple eigenvalues, the formulae above are still valid, but the projectors $\Pi_j^b$ above must be replaced by projectors upon eigenspaces. One can show that (\ref{Born21}) then gives a precise version of Born's rule for this case.

\begin{proof}
Look first at the case with unique eigenvalues. Then Born's rule says
\[P(\theta^b=u_j^b|\psi)=\langle\psi|b;j\rangle\langle b;j|\psi\rangle .\]
Let then the eigenvalues move towards coincidence. Let $C_k =\{j: u_j^b=v_k^b\}$. Then by continuity from the previous equation we get
\[P(\theta^b=v_k^b|\psi)=\sum_{j\in C_k}\langle\psi|b;j\rangle\langle b;j|\psi\rangle =\langle\psi |\varPi_k^b| \psi\rangle  = \| \varPi_k^b| \psi\rangle\|^2 .\]
\end{proof}

Note that in general $P(\theta^b=u_j^b\ \text{and\ then}\ \theta^c=u_i^c|\psi)\ne P(\theta^c=u_i^c$ and then $\theta^b=u_j^b|\psi)$. Measurements do not necessarily commute.

\section{Generalizations}

Using a suitable projection, the formula can be generalized to the case where also the accessible variables  $\theta^a$ is not necessarily maximal. There is also a variant for a mixed state involving $\theta^a$.

First, define the mixed state associated with any accessible variable $\theta$. We need the assumption that there exists a maximal accessible variable $\eta$ such that $\theta=f(\eta)$ and such that each distribition of $\eta$, given some $\theta=u$, is uniform. Furthermore some probability distribution of $\theta$ is assumed. Let $\Pi_u$ be the projection of the operator of $\theta$ upon the eigenspace associated with $\theta=u$. Then define the mixed state operator
\begin{equation}
\rho =\sum_j P(\theta=u_j)\Pi_{u_j}=\sum_i\sum_j P(\eta=v_i|\theta=u_j=f(v_i))P(\theta=u_j)|\psi_i\rangle\langle\psi_i |,
\label{Born2}
\end{equation}
where $|\psi_i\rangle$ is the state vector associated with the event $\eta=v_i$ for the maximal variable $\eta$.

From this, we can easily show from (\ref{Born1}) (assuming that the maximal $\eta^a$ corresponding to $\theta^a$ also is a function of $\phi$) that in general
\begin{equation}
P(\theta^b =v|\rho^a)= \mathrm{trace}(\rho^a\Pi_v^b),
\label{Born3}
\end{equation}
with an obvious meaning given to the projection $\Pi_v^b$.

An important observation is that this result is not necessarily associated with a microscopic situation. The result can also be generalized to continuous theoretical variables by first approximating them by discrete ones. For continuous variables, Born's formula is most easily stated on the form
\begin{equation}
E(\theta^b|\rho^a)=\mathrm{trace}(\rho^a A^{\theta^b}).
\label{Born}
\end{equation}

Note again that we in this formula do not assume that the accessible variable $\theta^b$ is maximal. Hence a corresponding formula is also valid for any function of $\theta^b$, for instance $\mathrm{exp}(i\theta^b x)$ for some fixed $x$. The operator corresponding to a function of $\theta^b$ can be found from the spectral theorem. From this, the probability distribution of $\theta^b$, given the information in $\rho^a$, can be recovered.

\bigskip

I can also generalize to the case where the final measurement is not necessarily perfect. Let us assume future data $z^b$ instead of a perfect theoretical variable $\theta^b$. Strictly speaking, for this case the focused likelihood principle is still valid under the following condition: $p(z^b |\theta^b =u_j)=p(z^b|\theta^b=u_k)$ implies $u_j=u_k$. This will not be needed here. We can define an operator corresponding to $z^b$ by
\begin{equation}
B^{z^b}=\sum_j p(z^b |\theta^b =u_j)\Pi_{u_j}^b,
\label{Born4}
\end{equation}
and, conditioning upon the events $\theta^b=u_j$ and following version (\ref{Born3}) of the Born formula, we obtain
\begin{equation}
p(z^b|\rho^a)=\mathrm{trace}(\rho^a B^{z^b}).
\label{Born5}
\end{equation}

\section{Intersubjectivity and QBism}

Consider two remote observers $O_1$ and $O_2$ who perform joint measurments on a system $\mathcal{S}$. Let their observations at time $t$ be $\theta_1(t)$ and $\theta_2(t)$, and let these correspond to operators $M_1(t)$ and $M_2(t)$. Khrennikov (2024) considers this situation, and assumes that $[M_1(t), M_2(t)]=0$. Is this possible?

In my terminology it is only possible if $\theta_1(t)$ and $\theta_2(t)$, in some sense can be given meaning at the same time, are both accessible. They can then not each be maximal, but one can imagine a situation where the vector $(\theta_1(t), \theta_2(t))$ is maximal. At least it has to be accessible to some agent. This agent can be a third observer, observing both $O_1$ and $O_2$.

Khrennikov then refers to a Theorem due to Ozawa (2019): Two observers performing the joint local and probabiliy reproducible measurements of the same observable $A$ on the system $\mathcal{S}$ should get the same outcome with probability 1. He says that this challenges QBism.

This last challenge is met by Schack (2023). His arguments are based on the quantum formalism, and the QBist interpretation of this formalism. I will not here go into his detailed mathematics, but only his interpretation of this mathematics. Here are two citations:

`The quantum formalism is a tool that any agent can use to optimize their choice of actions.'

`The quantum formalism does not describe nature in absence of agent, but instead is \emph{normative}, i.e., answers the question of how one \emph{should} act.'

I agree completely that quantum probabilities should be attached to an agent (or to a communicating group of agents), but here the agreement stops. First, I will allow any agent, not only one that is familiar with the quantum formalism. Next, I look upon quantum probabilities as descriptive, not normative. This is also my background for interpreting Ozawa's Theorem.

Here is a citation from Section 3 in Schack's article:

`As a mathematical result, Ozawa'stheorem says nothing about intersubjectivity or different observers. To arrive at their interpretation, both Ozawa and Khrennikov have to make the additional assumption that their scenario - two different observers interacting with a system followed by measurements on the meters - describes two different observers observers measuring the same system observable.'

He then goes on arguing that this assumption is incompatible with QBism. He says that from a QBist perspective, Ozawa's Theorem is about measurements that  \emph{a single agent}, say, Eve, contemplates performing on a system and two meters. The assumption that the theorem is about measurement results of two different observers violates QBism's key tenet that the quantum formalism should be viewed as a single-agent theory.

If this is the case, I disagree with the main basis of QBism. In my view, one can well imagine two observers $O_1$ and $O_2$ measuring the same system. But then it must be done in such a way that the vector of results $(\theta_1,\theta_2)$ is accessible to some agent. What does this mean? As I see it, it means that a third observer - you may well call her Eve - may be able to observe $O_1$ and $O_2$ during their measurements, and then able to record their results all the time. So one can well regard quantum mechanics as a single agent theory. From the point of Eve here, it can be taken to describe what she observes.

However, note that my basic mathematical theory (Helland, 2024 a,b) can be interpreted in two directions. It can be seen as a single agent theory, but it can also be seen as a theory of the joint minds of a group of communicating actors, where their communication involves the relevant theoretical variables.

Zwirn (2024), in his discussion of the Khrennikov/QBists articles, considers intersubjectivity from the point of view of Convivial Solipsism. ConSol describes the interaction of two observers that are not initially communicating. What one observer then learns about the other observer's values, are seen as a measurement. Inside ConSol, everything is relative to one unique observer. I can agree with this if the `unique observer' also includes groups of observers that initially have communicated on everything that is relevant. Each such `unique observer' will have his or her own perspectival reality. In this article I will not go into this in detail. I will only say that I in large terms agree with both ConSol and QBism when it concerns how an agent learns a new result, and then updates his or her own state. To me, a pure state always belongs to a `unique observer', and it can always be seen as the result of asking a maximal question to nature or to another agent, and then obtaining a definite answer. Depending on the situation, this answer may be known, or it can just be the expectation of an agent. I see the state concept as useful also in the last case.

So, answering a question raised by a referee, I agree with the QBist statement that different measurers may assign different (pure) states to a system. 

However, I see the whole of quantum theory as relative to a fixed inaccessible variable $\phi$, varying on a space $\Omega_\phi$. In simple physical systems, such a $\phi$ may easily be found, but, dependent on our philosophy, the general interpretation of $\phi$ may vary.

Taking for simplicity $\phi$ to be discrete, and assiming that for some higher being called God, $\phi$ takes some value $u$, this value describes a very abstract pure state $|\psi\rangle$, which could be called `the wave function of the world'. Personally, I see this concept as being less fruitful. Neither $u$ nor $|\psi\rangle$ could be observed by any human observer. I look upon $\phi$ as a variable, an inaccessible one, and the values that it may take, are just hypothetical.

\section{Interpretation and foundation of quantum mechanics}

Unfortunately, there are many different, mutually incompatible interpretations of quantum mechanics. The relevant Wikipedia article mentions 16 different interpretations. QBism is one of them. There is a large literature on QBism, some of it referred to in Schack's article. I agree of much that is written in this literature, but, as stated in the previuos sections, I disagree with their views on quantum probabilities.

The question of when one shall use classical probabilities and when one shall use quantum probabilities, is crucial; see Pothos and Busemeyer (2023) and references there, and from a statistical point of view, an example in Subsection 5.6.4 in Helland (2021). From a QBist point of view, this question is taken up in DeBrota et al. (2020a,b). The crucial argument is at the beginning of DeBrota et al. (2020a): Assume a POVM $\{D_j\}$ and a mixed quantum state $\rho$ belonging to some agent. Then the Born rule gives probabilities $Q(D_j)=\mathrm{tr}\rho D_j$ for the outcomes of the agent's measurement. The main concept of the papers are minimally complete POVMs (MICs). These sets of operators form a bases for the vector space of Hermitian operators and lead to probability distributions with fewest number of entries necessary for reconstructing the quantum state. 

The argument runs as follows: In addition to the actual measurements assume `measurements in the sky' $\{H_i\}$. Let $P(H_i)$ be their probabilities, and $P(D_j|H_i)$ the condition probabilities, given $H_i$ for subsequent measurements of $\{D_j\}$. Furthermore, let $\{\sigma_j\}$ be a basis in the state space, and define the matrix $\Phi$ by its inverse $[\Phi^{-1}]_{ij}=\mathrm{tr}H_i\sigma_j$. Then in op.cit the following basic quantum law in matrix notation is shown:
\begin{equation}
Q(D)=P(D|H)\Phi P(H).
\label{Phi}
\end{equation}
This is seen as a quantum version of the law of total probability. Note that while $P(H)$, $P(D|H)$, and $Q(D)$ are probabilities, $\Phi P(H)$ often is not.

This is the basis for studying MICs defined by $\{H_i\}$. In particular, the symmetric IC POVM (SIC) is a MIC for which all the $H_i$s are of rank 1. The existence of SICs of all dimensions is an open mathematical problem. In DeBrota et al. (2020b), the MICs are studied in detail, and it is shown how these MICs illuminate the structure of quantum theory and how it departs from the classical.

All this are interesting investigations. There is a SIC version of the Born rule, and this SIC solution is shown to be optimal in at least two different ways. But note that this all depends on the `measurements in the sky' giving $H_i$s as described above. One can also refer to DeBrota et al. (2021), where the QBist version of the Born rule is given explicitly, and where detailed arguments are given.

My claim is that this does not give the most general version of the Born rule. For this, I refer to the derivation given above, where no such $H_i$s are assumed. The QBist construction is connected to subjective beliefs, and the $P(H_i)$s are subjective probabilities.

To me, the basic concepts are not only belief or knowledge; it is that of decisions, decisions whose theory can be derived from a fundamental theory based on theoretical variables. You can decide to measure the spin of a silver atom in the $x$-direction or in the $z$-direction. As the result of a (precise) measurement you will get different state vectors in the two cases. To me, it is less important whether this is a state of belief or a state of knowledge. The main thing is that you have decided upon a measurement direction and then measured.

This way of thinking carries over to daily life situation. We go through life taking decision after decision, making choice after choice. Each decision is intuitive, it can be based upon our beliefs, or our knowledge, or both. It is interesting that quantum-like models now are beginning to be applied in decision-contexts and in related macroscopic contexts, say in psychology and economics; see the work of Andrei Khrennikov and collaborators, and also independent works in quantum decision theory. In Helland (2023a) I try to connect this to my basic approach to quantum foundation.

As I see it, any quantum interpretation should be coupled to a quantum foundation. My views on the quantum foundation are now described in Helland (2024a,b). This naturally leads to what I call a general epistemic interpretation of the theory. It is based upon theoretical variables that are connected to an agent or to a group of communicating agents in some fixed context. Some of these variables are accessible to the agent, others are inaccessible. My first main theorem states that in a situation with two different accessible variables that in this sense are maximal to the agent, there can be defined a Hilbert space $H$ such that all accessible variables are associated with self-adjoint operators in $H$. The eigenvalues of an operator $A$ coincide with the possible values of the associated variable. An accessible variable is maximal if and only if the associated operator has only one-dimensional eigenspaces.

In general, these results require some symmetry assumptions, but in the discrete case, it seems as if these symmetry assumptions can be dispensed with, see Helland (2024a). In the discrete case, a pure state may be identified with a question involving a maximal accessible variable together with a sharp answer to this question. Of course this variable may be a vector, implying several (commuting) partial questions.

In addition to these basic results, I need arguments for the Born formula and for the Schr\"{o}dinger equation. Both issues are addressed in Helland (2021). My assumptions behind several versions of Born's formula are given above. 

What are the prices payed for all this? First, some simple axioms are to be assumed. Most of them are rather obvious, but one should be mentioned: There exists an inaccessible variable $\phi$ such that all accessible variables are functions of $\phi$. In several physical examples, $\phi$ can easily be constructed. One can also discuss purely statistical applications. As a very general axiom, valid for all agents in all possible situations, one can take several points of view; one option is to argue for a religious perspective, see Helland (2023c).

A second price should be mentioned. The theory starts by constructing operators associated with all accessible variables. Pure state vectors are then only introduced as eigenvectors of some physically meaningful operator. This seems to impose a limitation on the superposition principle. This can be discussed, and should be discussed, but it should be remarked that this theory also includes some entangled state vectors (Helland, 2024b). On the good side, this version of the quantum theory leads to a simple understanding of so-called quantum paradoxes, like Schr\"{o}dinger's cat, the two-slit experiment and Wigner's friend, and there are links towards relativity theory and quantum field theory; see Helland (2023b) and Helland and Parthasarathy (2024).

\section{Conclusions}

`The discussion of quantum foundation and quantum interpretation will probably continue. I have presented my own views in several articles. This leads to a consistent theory, and a theory that also can be explained to outsiders. I see that as a great advantage. In particular, the theory can easily be explained in the discrete case, which has many applications, and is treated in very many textbooks. The continuous case can be approached by taking limits from a discrete construction; see again Helland (2021), but it is important also to have an independent basis for this case (Helland, 2024a,b).

The introduction of quantum probabilities requires extra assumptions as described above. One of these assumptions, the likelihood principle, is related to statistical theory. This opens for a possible communication between statisticians and quantum physicists, a communication that up to now has been very scarce. With the rapid progress now of artificial intelligence, which is closely connected to statistics (Hastie et al., 2017), and the fact that there now are appearing several articles connecting artificial intelligence to quantum mechanics, see Bharti et al. (2020), Dunjko and Briegel (2019), Dunjko et al. (2016), Rupp (2015), and Zhu et al. (2023) , such communication should be treated as being of some importance. It has been a major goal of my approach.

The other assumption behind the Born rule relates to the ideals of the relevant agent. These are assumed to be of a kind that can be modeled by an abstract or concrete higher being, considered by the agent to be perfectly rational. In Helland (2023a), such ideals are discussed in connection to decision processes.

My theory can be taken as a basis for reviewing discussions within the quantum community. In the present article I have considered the recent discussion between Khrennikov (2024) and a couple of QBists. From my point of view, I have stated some arguments against a pure QBist interpretation as the only solution. My own approach leads to a general epistimic interpretation, containing QBism as a special case.

For many years there has been discussions in the statistical society between Bayesians and frequentists, but these discussions have now calmed down. The best statisticians, see e.g. Efron (2015), have used tools from both schools. The discussions have also involved a third school, fiducial inference, founded by Fisher (1930, 1956), then discarded by most statisticians, but in recent years revived by statisticians like Hannig et al. (2016) and Taraldsen and Lindquist (2024). A forum for such discussions have been international BFF conferences (Bayesian, Frequentist, Fiducial, also informally named Best Friends Forever). 

In some future one could hope for extended discussions of this kind on the foundation of empirical science, involving scientists from several cultures, including quantum physicists and statisticians. This will require a common language, a language that has been sought for in my recent articles.

\section*{Acknowledgements}

I want to thank Gleb A. Skorobagatko and Thiago Prudenscio for their comments to a briefer version of this article. I am also grateful to two referees for their constructive remarks.

\section*{References}

Barndorff-Nielsen, O.E., Gill, R.D., and Jupp. P.B. (2003). On quantum statistical inference. \textit{J. Royal Statist. Soc., Series B: Statistical Methodology} \textbf{65} (4), 775-804.
\smallskip

Berger, J.~O., and Wolpert, R.~L. (1988). \newblock \textit{The likelihood principle.}\newblock Hayward, CA: Institute of Mathematical Statistics.
\smallskip

Bernardo, J.M. and Smith, A.F.M. (2009). \textit{Bayesian Theory.} Wiley.
\smallskip

Bharti, K., Haug, T., Vedral, V. and Kwek, L.-C. (2020). Machine learning meets quantum foundations: A brief survey. \textit{AVS Quantum Science} 2, 034101.
\smallskip

Busch, P. (2003). Quantum states and generalized observables: A simple proof of Gleason's Theorem. \textit{Phys. Rev. Letters} \textbf{9} (12), 120403.
\smallskip

Busch, P. (2003). Quantum states and generalized observables: A simple proof of Gleason's Theorem. \newblock \textit{ Physical Review Letters} \textbf{91} (12), 120403.
\smallskip

Busch, P., Lahti, P., Pellonp\"{a}\"{a}, J.-P., and Ylinen, K. (2016). \newblock\textit{Quantum measurement}. Berlin: Springer.
\smallskip

Campanella, M., Jon, D. and Mongiovi, M.S. (2020). Interpretative aspects of quantum mechanics. In \textit{Matteo Campanella's Mathematical Studies.} Springer, Cham.
\smallskip

Caves, C. ~M., Fuchs, C.~A. and Schack, R. (2002). Quantum probabilities as Bayesian probabilities. \newblock \textit{Physical Review,} 65, 022305.
\smallskip

DeBrota, J.B., Fuchs, C.A. and Stacey, B.C. (2020a). Symmetric informationally complete measurements Identify the irreducible difference between classical and quantum systems. \textit{Physical Review Research} 2, 013074.
\smallskip 

DeBrota, J.B., Fuchs, C.A. and Stacey, B.C. (2020b). The varieties of minimal tomographically complete measurements. arXiv: 1812.08762v5 [quant-ph].
\smallskip

DeBrota, J.B. Fuchs, C.A., Pienaar, J.L. and Stacey, B.C. (2021). Born's rule as a quantum extension of Bayesian coherence. \textit{Physical Review A} 104, 022207.
\smallskip

de Finetti, B. (1972). \textit{Probability, Induction and Statistics} Wiley, New York
\smallskip

Dunjko, V.  and Briegel, H.J. (2019). Machine learning \& artificial intelligence in the quantum domain: a review of recent progress. \textit{Rep. Prog. Phys.} \textbf{81}, 074001.
\smallskip

Dunjko, V., Taylor, J.M. and Briegel, H.J. (2016). Quantum-enhanced machine learning. \textit{Physical Review Letters} 117, 130501.
\smallskip

Efron, B. (2015). Frequentist accuracy of Bayesian estimates. \textit{Journal of the Royal Statistical Society B} 77, 617-646.
\smallskip

Fisher, R.A. (1930). Inverse Probability. \textit{ Proceedings of Cambridge Philosophical Society} 26, 528-535.
\smallskip

Fisher, R.A. (1956). \textit{Statistical Methods and Scientific Inference.} Hafner Press.
\smallskip

Good, I.J. (1983). \textit{Good Thinking: the Foundations of Probability and its Applications} University of Minnesota Press.
\smallskip

Hammond, P.~J. (2011). \emph{Laboratory games and quantum behavior. The normal form with a separable state space}. Working paper. Department of Economics, University of Warwick.
\smallskip

Hannig, J., Iyer, H., Lai, R.C.S. and Lee, T.C.M. (2016). Generalized fiducial inference: A review and new results. \textit{Journal of the American Statistical Association} 111(515), 1346-1361.
\smallskip

Hastie, T., Tibshirani, J. and Friedman, R. (2017). \textit{The Elements of Statistical Learning.} Springer, New York.
\smallskip

Haven, E. and Khennikov, A. (2013). \textit{Quantum Social Science.} Cambridge: Cambridge University Press.
\smallskip

Haven, E. and Khrennikov, A. (2016). Statistical and subjective interpretations of probability in quantum-like models in cognition and decision-making. \textit{J. Mathematical Psychology} 74, 82-91.
\smallskip

Helland, I.~S. (2006). Extended statistical modeling under symmetry; the link toward quantum mechanics. \newblock \textit{Annals of Statistics, 34}, 42--77.
\smallskip

Helland, I.~S. (2008). Quantum mechanics from focusing and symmetry. \newblock \textit{Foundations of Physics, 38}, 818--842.
\smallskip

Helland, I.~S. (2010). \newblock \textit{Steps Towards a Unified Basis for Scientific Models and Methods.} \newblock Singapore: World Scientific.
\smallskip

Helland, I.S. (2019). An epistemic interpretation and foundation of quantum theory. arXiv:1905.06592 [quant-ph].
\smallskip

Helland, I.S, (2021).
\textit{Epistemic Processes. A Basis for Statistics and Quantum Theory.} 2. edition. Springer Nature Switzerland.
\smallskip

Helland, I.S, (2022a). On reconstructing parts of quantum theory from two related maximal conceptual variables.  \textit{International Journal of Theoretical Physics} 61, 69.
\smallskip

Helland, I.S. (2022b). The Bell experiment and the limitation of actors. \textit{Foundations of Physics} 52, 55.
\smallskip

Helland, I.S. (2023a). On the foundation of quantum decision theory. arXiv:2310.12762 [quant-ph]
\smallskip

Helland, I.S. (2023b). Possible connections between relativity theory and a version of quantum theory based upon theoretical variables. arXiv: 2305.15435 [physics.hist-ph]
\smallskip

Helland, I.S. (2023c). A simple quantum model linked to decisions. \textit{Foundations of Physics} 53, 12.
\smallskip

Helland, I.S. (2023d). On the Bell experiment and quantum foundation. \textit{Journal of Modern and Applied Physics} 6 (2), 1-5. arXiv: 23305.05299 [quant-ph].
\smallskip

Helland, I.S, (2024a).
An alternative foundation of quantum theory. \textit{Foundations of Physics} 54, 3.
\smallskip

Helland, I.S. (2024b). A new approach towards quantum foundation and some consequences. arXiv:2403.09224 [quant-ph].
\smallskip

Helland, I.S. and Parthasarathy, H. (2024). \textit{Theoretical Variables, Quantum Theory, Relativistic Quantum Field Theory, and Quantum Gravity.} Manakin Press, New Dehli.
\smallskip

Holevo, A.~S. (2001). \newblock \textit{Statistical structure of quantum theory.} Springer, Berlin.
\smallskip

Khrennikov, A. (2023). \textit{Open Quantum Systems in Biology, Cognitive and Social Sciences.} Springer Nature Switzerland.
\smallskip

Khrennikov, A. (2024). Ozawa's intersubjectivy theorem as objection to QBism individual agent perspective. \textit{Internat. J. Theor. Phys.} \textbf{63}, 23.
\smallskip

Lehmann, E.~ L. and Casella, G. (1998). \textit{Theory of point estimation.} Springer, New York

Ozawa, M. (2019). Soundness and completeness of quantum root-mean-square errors. \textit{Quant. Inf.} 5,1.
\smallskip

Ozawa, M. and Khrennikov, A. (2021). Modeling combination of question order effect, response replicability, and QQ-equality with quantum instruments. \textit{J. Mathematical Psychology} 100, 102491.
\smallskip

Pothos, E.~M. \& Busemeyer. J.~R: (2022). Quantum Cognition. \newblock \textit{Annual Review of Psychology} 73, 749-778.
\smallskip

Rupp, M. (2015). Machine learning for quantum mechanics in a nutshell.\textit{International Journal of Quantum Chemistry} 115, 1058-1073.
\smallskip

Savage, L.J. (1972). \textit{The Foundations of Statistics.} Dover, New York.
\smallskip

Schack, R. (2023). When will two agents agree on quantum mesurement outcome? Intersubjective agreement in QBism. arXiv: 2312.07728v1 [quant-ph]
\smallskip

Schweder, T. and Hjort, N.~L. (2002). Confidence and likelihood. \textit{Scandinavian Journal of Statistics} 28, 309--332.
\smallskip

Shrapnel; S., Costa, F. \& Milburn, G. (2017). Updating the Born rule. arXiv: 1702.01845v1 [quant-ph].
\smallskip

Stacey, B. C. (2023). Whose probabilities? About what? A reply to Khrennikov. arXiv: 2302.09475v1 [quant-ph]
\smallskip

Taraldsen, G. and Lindquist, B.H. (2024). Fiducial inference and decision theory. In: Berger, J., Meng, X.-L., Reid, N. and Xie, M. [Ed.] \textit{Handbook of Bayesian, Fiducial, and Frequentist Inference} Chapman and Hall. ArXiv: 2112.07060 [stat.ME].
\smallskip

Veloz, T., Khrennikov, A., Toni, B. and Castillo, R.D. [ed.] (2023). \textit{Trends and Challenges in Cognitive Modeling.} Springer Nature Switzerland. 
\smallskip

von Mises, R. (1981). \textit{Probability, Statistics, and Truth} Dover, New York
\smallskip

von Neumann, J. (1927). Wahrscheinlichkeitstheoretischer Aufbau der Quantenmechanik. \newblock \textit{Nachrichten von der Gesellschaft der Wissenschaften zu G\"{o}ttingen, Mathematisch-Physikalische Klasse, 1927}, 245--272.
\smallskip

Zhu, Y., Wu, Y.-D., Liu, Q., W., Y. and Chiribella, G. (2023). Predictive modelling of quantum process with neural networks. arXiv: 2308.08815 [quant-ph].
\smallskip

Zwirn, H. (2024). Is intersubjectivity proven? A reply to Khrennikov and to QBists. arXiv:24054.04367 [quant-ph]

\newpage

\section*{Appendix 1}
\underline{Proof of the Direct Part of Proposition 1.}

Define $b_k=p(z^b|\tau,\theta^b=u_k^b)$ and $c_k= p(z^c|\tau,\theta^c=u_k^c)$. Then (\ref{effect}) is
\[ \sum_k b_k |b;k\rangle\langle b;k|=\sum_k c_k |c;k\rangle\langle c;k|.\]
In these sums we can collect together terms with equal coefficients. Let $P_j=\sum_{k\in C_j}|b;k\rangle\langle b;k|$, where $C_j$ is defined such that $b_k=b_{k^{\prime}}$ when $k,k^{\prime}\in C_j$, similarly define $Q_j$ on the righthand side. Redefine $b_j$ as $b_{k_j}$ whenever $k_j\in C_j$, and redefine $c_j$ similarly. Then
\begin{equation}
\sum_j b_j P_j=\sum_j c_j Q_j,
\label{PQ}
\end{equation}
$\{P_j\}$ and $\{Q_J\}$ are orthogonal sets of projection operators, $b_j\ne b_{j^{\prime}}$ whenever $j\ne j^{\prime}$ and $c_j\ne c_{j^{\prime}}$ whenever $j\ne j^{\prime}$. We can order the terms in (\ref{PQ}) such that $b_1>b_2>\ldots>0$ and $c_1>c_2>\ldots>0$. Furthermore we can multiply each side of (\ref{PQ}) with itself any number of times, giving
\[\sum_j b_j^m P_j=\sum_j c_j^m Q_j\]
for $m=1,2,\ldots$. When $m$ is large enough, the first term on each side of this equation will dominate completely, and we must have $b_1^m P_1=c_1^m Q_1$ for all large enough $m$. But since $P_1$ and $Q_1$ are projection operators, this is only possible if $b_1=c_1$ and $P_1=Q_1$.

Then we can subtract the term $b_1 P_1$ from the lefthand side of (\ref{PQ}), subtract the equal term $c_1 Q_1$ from the righthand side of the equation, and repeat the argument. It follows that $b_j=c_j$ and $P_j=Q_j$ for each $j$, which is the conclusion of the proposition.
\qed

\newpage

\section*{Appendix 2}
\underline{Proof of Busch's Theorem for~the~Finite-Dimensional Case}

The main point of the proof is to show that any generalized probability measure on effects extends to a unique positive linear functional on the vector space of all bounded linear Hermitian operators. This is done
in steps.

\begin{enumerate}
\item[1.] It is trivial that $\mu(E)=n\mu(\frac{1}{n}E)$ for all positive integers. It follows that $\mu(pE)=p\mu(E)$ for all rational numbers in $[0,1]$. By approximating from below and from above by rational numbers, this
implies that $\mu(\alpha E)=\alpha\mu(E)$ for all real numbers $\alpha$ in $[0,1]$.

\item[2.] Let $A$ be any positive bounded operator in $H$. Then there is a positive number $\alpha$ such that $\langle u|Au\rangle \le \alpha$ for all unit vectors $u$. Then $E$ defined by $E=(1/\alpha)A$ is an effect.
Thus we can always write $A=\alpha E$ for an effect $E$. Assume now that there are two effects $E_1$ and $E_2$ such that $A=\alpha_1 E_1 =\alpha_2 E_2$. Assume without loss of generality that $\alpha_2 >\alpha_1 >0$.
Then by (1) $\mu(E_2)=\frac{\alpha_1}{\alpha_2}\mu(E_1)$, so $\alpha_1 \mu(E_1)=\alpha_2 \mu(E_2)$. Therefore we can uniquely define $\mu(A)=\alpha_1 \mu(E_1)$.

\item[3.] Let $A$ and $B$ be positive bounded operators. Take $\gamma>1$ such that $\frac{1}{\gamma} (A+B)$ is an effect. Then we can write $\mu(A+B)$ as $\gamma\mu(\frac{1}{\gamma}(A+B))=\gamma\mu(\frac{1}{\gamma}A)
+\gamma\mu(\frac{1}{\gamma}B)=\mu(A)+\mu(B)$.

\item[4.] Let $C$ be an arbitrary bounded Hermitian operator. Assume that we have two different decompositions $C=A-B=A^{\prime}-B^{\prime}$ into a difference of positive operators. Then $A+B^{\prime}=A^{\prime}+B$ implies $\mu(A)+\mu(B^{\prime})=\mu(A^{\prime})+\mu(B)$.
Hence $\mu(A)-\mu(B)=\mu(A^{\prime})-\mu(B^{\prime})$, so we can uniquely define $\mu(C)$ as $\mu(A)-\mu(B)$. It follows then easily from (3) that $\mu(C+D)=\mu(C)+\mu(D)$ for bounded Hermitian operators.

\item[5.] This is extended directly to $\mu(C_1+\ldots+C_r)=\mu(C_1+\ldots+C_{r-1})+\mu(C_r)=\mu(C_1)+\ldots+\mu(C_r)$ for finite sums.
\end{enumerate}

Let $\{|k\rangle ;k=1,\ldots,n\}$ be a basis for $H$. Then for any Hermitian operator $C$ we can write $C=\sum_{i,j} c_{ij}|i\rangle\langle j|$, where $c_{ij}$ are complex numbers satisfying $c_{ij}*=c_{ji}$. Define the operator $\rho$
by $\rho_{ij}=\mu(|i\rangle\langle j|)$. Then $\rho$ is a positive operator since $\langle v|\rho v\rangle =\mu(|v\rangle \langle v|)$ for any vector $|v\rangle$. Also
\[\mathrm{trace}(\rho)=\sum_i \rho_{ii}=\sum_i \mu(|i\rangle\langle i|)=\mu(\sum_i |i\rangle\langle i|)=\mu(I)=1,\]
so $\rho$ is a density operator.

We have $\mu(C)=\sum_{i,j}\rho_{ij}c_{ij}=\mathrm{trace}(\rho C)$, and this holds in particular when $C$ is an effect.

\end{document}